%% file: main.tex
\documentclass[
    aps,prl,twocolumn,
	groupedaddress,superscriptaddress,
	amsfonts,amssymb,amsmath,
	citeautoscript,longbibliography,
	letterpaper, nofootinbib
	]{revtex4-2}

\input{main-setup}

\graphicspath{{figures/}}
\setboolean{togglecomments}{true}
\setboolean{toggletodos}{true}
\setboolean{togglechanges}{false} 

\begin{document}
\title{Three-dimensional confinement of light in photonic crystals without bandgaps}

\author{Manxi Shi$^\bigstar$}
\affiliation{Department of Physics, Massachusetts Institute of Technology, Cambridge, Massachusetts 02139, USA}

\author{Sachin Vaidya$^\bigstar$}
\email{svaidya1@mit.edu}
\affiliation{Department of Physics, Massachusetts Institute of Technology, Cambridge, Massachusetts 02139, USA}
\affiliation{Research Laboratory of Electronics, Massachusetts Institute of Technology, Cambridge, Massachusetts 02139, USA}

\author{Ali Ghorashi}
\affiliation{Department of Physics, Massachusetts Institute of Technology, Cambridge, Massachusetts 02139, USA}

\author{Steven G. Johnson}
\affiliation{Department of Mathematics, Massachusetts Institute of Technology, Cambridge, Massachusetts 02139, USA\\
$^\bigstar$ denotes equal contribution}

\author{Marin Solja\v{c}i\'{c}}
\affiliation{Department of Physics, Massachusetts Institute of Technology, Cambridge, Massachusetts 02139, USA}
\affiliation{Research Laboratory of Electronics, Massachusetts Institute of Technology, Cambridge, Massachusetts 02139, USA}

\begin{abstract}
We demonstrate that confinement of light in three dimensions is possible in photonic crystals without a complete photonic bandgap. Our approach exploits symmetry-protected quadratic degeneracies in the bulk band structure, where the photonic density of states vanishes at an isolated frequency. By introducing a point defect, we create a localized mode whose symmetry representation is incompatible with that of the surrounding bulk modes, suppressing coupling to propagating channels. The combination of vanishing density of states and a symmetry mismatch yields bound defect modes despite the absence of a spectral gap, as confirmed by time- and frequency-domain numerical simulations. This approach highlights the role of engineering the photonic environment around the defect to enable confinement, potentially providing a new route for designing optical cavities in three-dimensional photonic crystals.

\end{abstract}
\maketitle 
Photonic crystals enable control over the propagation and confinement of light through periodic dielectric structures \cite{Joannopoulos2008, johnson1999, john1987strong, yablonovitch1987inhibited, yablonovitch1993photonic}. Over the past two decades, substantial progress in nanofabrication has enabled the realization of three-dimensional photonic crystals across a wide range of platforms, including layer-by-layer lithographic assembly~\cite{Subramania2004, qi2004three}, direct laser writing via two-photon polymerization~\cite{deubel2004direct, vaidya2020, jorg2022observation, schulz2021, chernow2021, peng2016, chen2019observation}, colloidal self-assembly~\cite{wijnhoven1998preparation, lee2024dna, posnjak2024diamond, he2020colloidal}, implosion fabrication~\cite{salamin2026three, oran20183d}, and interference lithography~\cite{moon2006fabricating, divliansky2003fabrication}. These approaches support implementation across diverse materials such as semiconductors (e.g., silicon, III–V compounds), chalcogenides, polymers, and metals, significantly expanding the experimental accessibility of 3D photonic structures. 

Despite this progress, achieving strong three-dimensional confinement of light in such systems has remained challenging. Conventionally, confinement relies on complete photonic bandgaps, where propagation is forbidden within a frequency range~\cite{Joannopoulos2008}. Realizing such gaps in photonic crystals generally requires carefully engineered geometries, often obtained through a combination of physical intuition, iterative design, and, more recently, inverse-design and nonlinear topology optimization methods~\cite{men2014robust, cerjan2018complete, kao2005maximizing, kim2023automated}. These designs can be difficult to realize experimentally due to fabrication constraints and geometric complexity, limiting their practical applicability.

These limitations have spurred the exploration of other confinement mechanisms for light trapping in photonic devices. Several of these mechanisms—most notably index guiding, plasmonic confinement, and certain classes of bound states in the continuum—fundamentally rely on momentum conservation. This results in confined modes having at least one extended dimension that is either uniform or periodic. In index guiding, confinement arises from generalized total internal reflection, which requires that in-plane momentum be conserved across an interface, preventing coupling to radiative modes. Similarly, plasmonic confinement relies on surface-bound modes whose large in-plane momentum lies outside the light cone, inhibiting radiation into free space~\cite{jablan2009plasmonics, khurgin2015ultimate, yang2017low}. Bound states in the continuum exploit symmetry and interference to enforce decoupling from radiative channels, but this protection is often (but not always) contingent on momentum conservation~\cite{monticone2018trapping, vaidya2021point, hsu2013, gao2019bound, kang2023applications, cerjan2019bound, cerjan2021observation, watts2002electromagnetic, xu2004modal}.

Anderson localization provides a route to confinement that does not rely on momentum conservation and can, in principle, give rise to full three-dimensional confinement~\cite{anderson1958absence, john1987strong, segev2013anderson, yamilov2023anderson}. It has been successfully employed for random lasing and cavity QED, where multiple scattering in disordered media is advantageous~\cite{liu2014random, sapienza2010cavity}. However, this mechanism lacks deterministic control: the spatial location, frequency, and quality factor of localized states are highly sensitive to disorder realizations. Moreover, recent studies have argued that longitudinal electric fields prevent such a disorder-induced localization of light in three-dimensional dielectric media~\cite{yamilov2023anderson, yamilov2025anderson}. More broadly, the extent to which full three-dimensional confinement can be achieved in ordered, all-dielectric systems without relying on complete photonic bandgaps remains an active question.

In this work, we uncover a mechanism for light confinement in three-dimensional photonic crystals (3D PhCs) without a complete bandgap. Our approach starts with a PhC with a symmetry-protected quadratic degeneracy in the bulk band structure, where the photonic density of states vanishes at an isolated frequency. By carefully introducing a point defect into the structure, we show that it is possible to generate a cavity mode whose symmetry representation is incompatible with that of the bulk modes of the PhC, thereby suppressing coupling to propagating channels. As a result, we obtain a strongly confined bound mode in fully three-dimensional, gapless PhCs, as confirmed by numerical simulations. Crucially, this form of confinement does not rely on momentum conservation or disorder, and represents a novel example of a bound state in the continuum. Finally, we identify all space groups that could potentially support this mechanism, expanding the design space of PhC cavities beyond conventional bandgap-based approaches.

\begin{figure}
    \centering
\includegraphics[width=\linewidth]{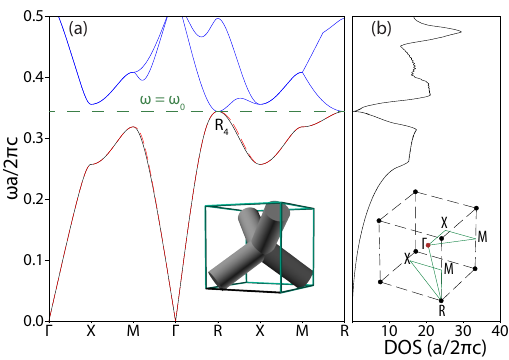}
    \caption{\textbf{Bulk band structure and density of states:} (a) The band structure of the 3D PhC is shown. The frequency of the symmetry-enforced and frequency-isolated threefold degeneracy at the R point is indicated by a green dashed line and labeled with its associated three-dimensional irrep. The inset shows the unit cell of the structure, consisting of four dielectric rods ($\varepsilon = 11$, $r_c/a = 0.18$, where $a$ is the lattice constant.) along the body diagonals of the cubic unit cell. (b) The associated density of states (DOS) of the photonic crystal. The DOS vanishes exactly at the frequency of the threefold degeneracy. The inset shows the path in the Brillouin zone over which the band structure was calculated. }
    \label{fig:figure1}
\end{figure}

We begin by considering a 3D PhC based on a simple cubic lattice of lattice constant $a$ with a unit cell containing four dielectric rods ($\varepsilon = 11$) oriented along the body diagonals [inset of Fig.~\ref{fig:figure1}(a)]~\cite{yang2017weyl}. This structure belongs to the  space group \#224 ($Pn\bar{3}m$), a nonsymmorphic space group that hosts three-dimensional irreducible representations (irreps) at certain high-symmetry points in the Brillouin zone, including at the $R$ point located at $(\pi/a, \pi/a, \pi/a)$~\footnote{We note that space group \#224 contains inversion. However, the unit cell shown in Fig.~\ref{fig:figure1}a is not centered at the inversion center, which is relevant for subsequent discussions. A small perturbation that only breaks inversion reduces the space group to \#208 ($P4_232$).}. At this point, the band structure exhibits a quadratic degeneracy involving three bands. By adjusting geometric parameters, specifically, the radius of the cylindrical rods $r_c/a$ and dielectric constant, we isolate this triply degenerate point from nearby bands [Fig.~\ref{fig:figure1}(a)]. Around this degeneracy, the photonic dispersion takes the form $\omega(k) \approx \omega_0 + \alpha |k - R|^2$, and the local DOS scales as $\sqrt{|\omega - \omega_0|}$~\cite{kittel2018introduction}, vanishing exactly at $\omega_0a/2\pi c \approx 0.345$ [Fig.~\ref{fig:figure1}(b)].

The central idea is to introduce a localized point defect into this structure whose geometry is tuned such that one of its resonant modes lies at $\omega_0$, coinciding with the bulk quadratic degeneracy. At this frequency, the vanishing DOS suppresses coupling to extended modes, resulting in partial confinement. However, this mechanism alone generally yields only weak localization~\cite{dirac1, diracexp1, diracexp2, diracexp3, medina2023corner, garcia2019tunable, ying2019extended, wehling2007local}. To achieve a bound mode at the defect, we additionally enforce a symmetry mismatch between the defect mode and the bulk modes. Specifically, the defect mode at $\omega_0$ is chosen such that it transforms under an irrep that is incompatible with the three-dimensional irrep of the bulk modes at the $R$ point, thereby prohibiting coupling by symmetry. 

As argued subsequently, the combined effect of DOS suppression and symmetry-enforced decoupling yields strongly localized, bound defect modes despite the absence of a complete photonic bandgap. From the perspective of spectral theory, such a bound mode may be viewed as a type of bound state in the continuum. This is because, although the density of states vanishes at $\omega_0$, this frequency remains part of the essential spectrum of the Maxwell operator describing the system $(\nabla \times \varepsilon^{-1}(\mathbf{r}) \nabla \times)$, and the defect mode corresponds to a normalizable state embedded within that continuum spectrum~\cite{teschl2014mathematical}.

\begin{figure}
\centering
      \includegraphics[scale=1]{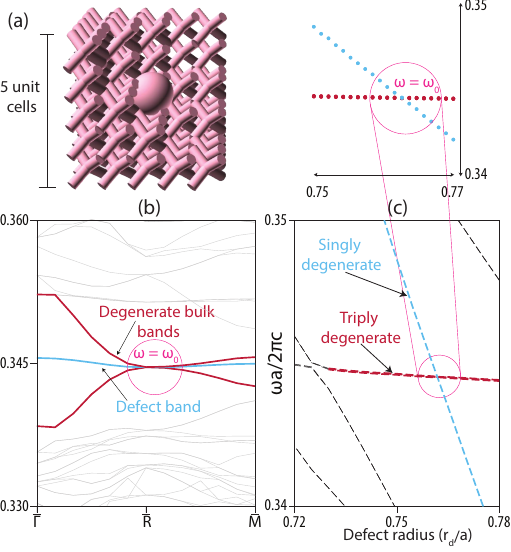}
  \caption{\textbf{Embedding a defect mode at the bulk degeneracy:} (a) A $5\times5\times 5$ supercell with a spherical defect introduced at the origin. One octant has been removed to show the defect at the center. (b) Band structure of this supercell with periodic boundaries in all three directions. The defect band (which transforms as a one-dimensional irrep, shown in light blue) can be tuned to be degenerate with three bulk bands (which transform as a three-dimensional irrep) at the $\bar{R}$ point. (c) Bands of this supercell at the $\bar{R}$ point as the defect radius, $\text{r}_d/\text{a}$, is varied. To show the lack of an avoided crossing as the defect and bulk bands become degenerate, we additionally zoom in on the range $0.75<r_d/a<0.77$, which is shown on top. }
  \label{fig:figure2}
\end{figure}

To motivate the mechanism of confinement, we first perform supercell calculations using the plane wave expansion (PWE) method as implemented in the MIT Photonic Bands (MPB)~\cite{MPB}. We introduce a solid spherical defect, made out of the same material as the rods, in a $5\times 5\times 5$ supercell (with periodic boundary conditions in all directions) at the centering shown in the inset of Fig.~\ref{fig:figure1}(a) [Fig.~\ref{fig:figure2}(a)]. Introducing this defect breaks the nonsymmorphic symmetries of the original space group \#224 (\#208 with the considered centering), as well as inversion, reducing the supercell structure to space group \#195 ($P23$). Importantly, this reduced space group continues to support three-dimensional irreducible representations at the $\bar{R}$ point of the smaller supercell Brillouin zone. As a result, the triply degenerate quadratic band touching of the bulk is preserved even in the presence of the defect [Fig.~\ref{fig:figure2}(b)].

It is therefore natural to view the supercell as a new periodic structure, into which the defect is incorporated. Within this enlarged unit cell, the original bulk modes form dispersive bands that retain their symmetry character at the $\bar{R}$ point, while the defect gives rise to additional nearly flat bands corresponding to spatially localized resonances. Crucially, some of these defect modes are singly degenerate and transform under a one-dimensional irrep of the reduced space group. By tuning the geometry of the defect, specifically the radius ($r_d/a$), we can force one such defect mode into a degeneracy with the triply degenerate bulk modes at the $\bar{R}$ point. Because the defect mode and the bulk modes transform under different irreps, symmetry forbids hybridization between them, as seen in the supercell band structure in Fig.~\ref{fig:figure2}(b). The absence of level repulsion is a direct signature of symmetry incompatibility of the modes. This feature can also be seen in the parameter scan of $r_d/a$ in Fig.~\ref{fig:figure2}(c).

\begin{figure}
  \centering
      \includegraphics[scale=1]{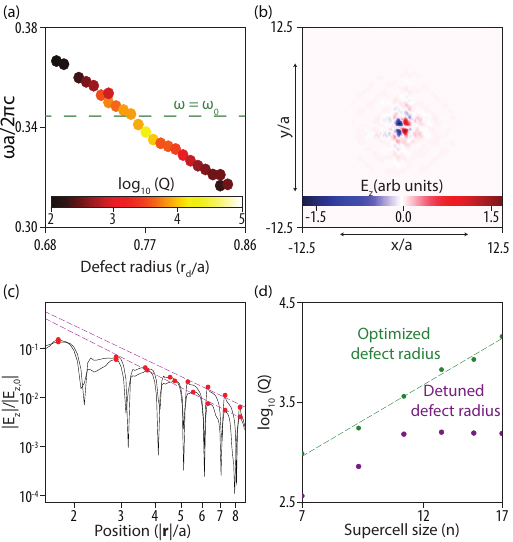}
    \caption{\textbf{Confinement and field profiles of the defect mode: } (a) The resonant frequency ($\omega$) and quality factor ($Q$) of the defect mode, obtained from FDTD, as a function of the radius of the spherical defect ($r_d/a$). $\omega_0$ marks the location of the bulk threefold degeneracy with vanishing DOS. This matches the dependence found from simulations using the PWE method in Fig.~\ref{fig:figure2}(c). (b) The $z$-component of the electric field ($E_z$) of defect mode at $\omega = \omega_0$, showing strong confinement. (c) A one-dimensional slice (at $y = z = 0$) across the confined defect mode showing algebraic decay of the field away from the defect. Dashed lines are linear fits (to the outermost peaks), corresponding to a decay of the electric field $E_z\sim |\mathbf{r}/a|^{-2.61}$ and $E_z\sim |\mathbf{r}/a|^{-2.78}$. (d) The scaling of $Q$ of the defect mode as a function of the system size ($n$) with frequency $\omega a/2\pi c= 0.345$ (green points) and a detuned frequency $\omega a/2\pi c = 0.327$ (purple points). The green line is a linear fit on a log-log scale, corresponding to a scaling of the quality factor, $Q\sim n^{3.06}$.}
  \label{fig:figure3}
\end{figure}

It is important to note that within this supercell description, the triply degenerate bulk bands remain frequency-isolated at the frequency $\omega_0$ for any supercell size. Furthermore, the bulk modes continue to have a well-defined symmetry content at $\omega_0$, while the tuned defect mode remains symmetry-decoupled from the bulk modes at exactly $\omega_0$. Therefore, the arguments made above extend to arbitrarily large supercells with a single spherical defect at the origin.

To validate our theoretical approach and to study the nature of confinement of this defect mode, we next perform finite-difference time-domain (FDTD) calculations as implemented in MEEP~\cite{oskooi2010meep}. The simulation domain consists of a cubic supercell of the photonic crystal and the spherical defect, with thick absorbing boundaries to eliminate reflections. We initialize the system with randomly distributed dipole sources with short-pulse excitations localized within and outside of the defect at the center. We analyze the resulting time-domain electromagnetic field using harminv, a tool that performs harmonic inversion on temporally decaying waveforms. Harminv extracts the complex eigenfrequencies of the system, allowing the determination of resonant frequencies [$\text{Re}(\omega)$] and corresponding $Q$-factors [$Q = -\text{Re}(\omega)/2\text{Im}(\omega)$] of the defect modes, even in the presence of overlapping or closely spaced resonances from the surrounding gapless structure~\cite{mandelshtam1997harmonic}. Modes with energy density concentrated at the defect center and high $Q$ factors are chosen for further analysis.

\begin{figure}[h]
    \centering
    \includegraphics[width=0.95\columnwidth]{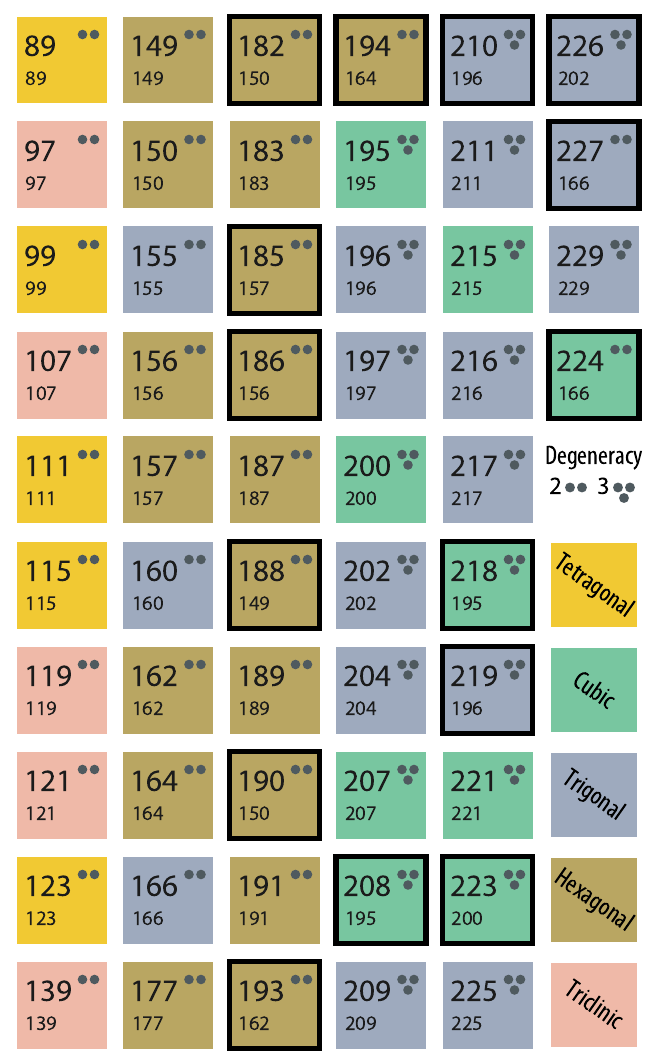}
    \caption{\textbf{Candidate symmetry settings for the localization mechanism discussed in this work.} A list of all space groups, organized by crystal system, that maintain multidimensional irreps in the presence of a spherical defect at the origin, even in the absence of time reversal symmetry. The smaller numbers in each box indicate the subgroups that the structure is reduced to upon introducing the defect (black outlines indicate cases when these differ). The dots indicate whether the bulk degeneracy induced by the multidimensional irrep is twofold or threefold. The case considered in this work corresponds to $(\#224)$ $\#208 \rightarrow\#195$.}
    \label{fig:figure4}
\end{figure}

Fig.~\ref{fig:figure3}(a) shows the complex frequency of the defect mode as a function of defect radius ($r_d/a$). As seen here, when the defect mode frequency coincides with that of the bulk threefold degeneracy, $\omega_0$, the $Q$-factor is maximized, reaching values approaching $10^5$ in modest system sizes~\footnote{We note that the small discrepancy (of $\sim1.5\%$) between the real parts of the frequencies of the defect mode from the FDTD method and the bulk calculations from the PWE method arises from differences in the two numerical methods.}. The spatial electric field distribution indicates strong localization around the defect [Fig.~\ref{fig:figure3}(b)]. A field profile of the mode along one direction (here, the $x$-direction) shows that the mode decays algebraically from the defect with distance $r$ approximately as $r^{-5/2}$ [Fig.~\ref{fig:figure3}(c)], as obtained from numerical fits. Due to the exponent being smaller than $-3/2$, the mode remains normalizable in three dimensions and therefore constitutes a genuinely bound state of this three-dimensional photonic system. Interestingly, unlike conventional exponentially localized cavity modes characterized by a finite localization length, this bound mode exhibits no intrinsic length scale due to its algebraic localization. 

Further evidence for the bound nature of the defect mode is provided by the scaling of the quality factor $Q$ with system size. For a normalizable mode with an algebraic tail of $\sim r^{-5/2}$, the field amplitude $|\mathbf{E}(L)|$ at the outer boundary of a finite system of linear size $L$ scales as $L^{-5/2}$, and the corresponding radiative flux through the boundary scales as $L^2 |\mathbf{E}(L)|^2 \sim L^{-3}$. Since the total energy stored in the mode approaches a finite value as $L \rightarrow \infty$, the radiative decay rate scales as $L^{-3}$, implying a quality factor scaling of $Q \sim L^3$. As shown in the simulations in Fig.~\ref{fig:figure3}(d), the quality factor exhibits this expected cubic scaling with system size when the defect mode is tuned to the confinement frequency $\omega_0$. This behavior is qualitatively distinct from that of an extended resonance, whose quality factor saturates with increasing system size due to its non-normalizable character. As seen in Fig.~\ref{fig:figure3}(d), detuning the defect radius turns the defect mode into a resonance, for which the growth of $Q$ with system size saturates.

We next present an argument for the observed algebraic localization of the bound mode that also elucidates the role of symmetry. The electric field of the defect mode satisfies the Lippmann--Schwinger equation $ \mathbf E(\mathbf r) = \int d^3r'\, \mathbf G(\mathbf r,\mathbf r';\omega_0)\, \Delta\varepsilon(\mathbf r')\, \mathbf E(\mathbf r')$, where $\mathbf G$ is the dyadic Green function of the unperturbed photonic crystal evaluated at the frequency $\omega_0$ of the isolated quadratic degeneracy. Expanding $\mathbf G$ in terms of the bulk Bloch eigenmodes near the degeneracy point, its long-distance behavior is governed by $\int d^3q\, e^{i\mathbf q\cdot\mathbf r}/{q^2} \propto r^{-1}$, implying that the bulk Green function generically contains a leading $r^{-1}$ contribution. Consequently, a defect mode would ordinarily inherit this $r^{-1}$ far-field tail through its coupling to the bulk modes, leading to a non-normalizable extended mode.

To determine the coefficient of this leading contribution, consider the overlaps $c_\alpha = \int d^3r\, \mathbf E_{\alpha,\mathbf k_0}^{*}(\mathbf r)\, \cdot \Delta\varepsilon(\mathbf r)\, \mathbf E_{\mathrm d}(\mathbf r)$, between the defect mode $\mathbf E_{\mathrm d}$ and the three bulk modes at the quadratic degeneracy labelled by $\alpha$. The amplitude of the asymptotic $r^{-1}$ tail is proportional to these overlap coefficients. However, in our construction, the representations of the bulk and defect modes are incompatible by symmetry, and therefore $c_\alpha = 0$. As a result, the coefficient of the leading $r^{-1}$ contribution vanishes identically and the defect mode cannot inherit the generic Green-function tail associated with the quadratic band touching. Its asymptotic decay must instead be at least as fast as $r^{-2}$, which results in a normalizable confined mode in three-dimensions and is consistent with our numerical results above. 

Although our results were obtained in the context of a specific geometry, we emphasize that the same localization mechanism may be engendered in a variety of space group settings. To support this, we find all symmetry settings for which our particular method of light localization could work. In particular, we find all space groups that can support both a multidimensional irrep (corresponding to degenerate extended bulk modes) and a one-dimensional irrep (corresponding to a localized defect mode) when modified with a spherical defect at the origin. This is equivalent to finding all space groups that, when stripped of their nonsymmorphic symmetries, yield subgroups with multidimensional irreps\footnote{This can be easily seen from the fact that the presence of a spherical defect at the origin only breaks nonsymmorphic symmetries.}. For brevity, in \cref{fig:figure4} we tabulate the space groups that have this property even in the absence of time reversal symmetry, so that the requirements are purely a consequence of the lattice geometry.

We note that these arguments disregard questions of ``energetics"--namely, whether the multidimensional irrep can be isolated in frequency and whether the defect mode may be tuned to be degenerate with the multidimensional bulk irrep (while maintaining spherical symmetry of the defect). Thus, the considerations above represent necessary but not sufficient conditions for enabling this confinement mechanism and serve as a starting point for inverse design methods.

To conclude, we have demonstrated that strong three-dimensional confinement of light can be achieved in photonic crystals without a complete photonic bandgap. Our approach involves a symmetry-protected quadratic degeneracy at a frequency where the photonic density of states of the environment vanishes, combined with a judiciously engineered defect whose symmetry representation is incompatible with that of the bulk modes. This interplay suppresses coupling to propagating channels and yields a strongly localized defect mode, which is an examples of a bound state in the continuum, as confirmed with numerical simulations. Furthermore, by identifying a broad set of space groups that may support the necessary symmetry structure, we show that this mechanism could represent a general design principle for photonic crystal nanocavities.

Looking forward, several directions emerge from this work. First, an important challenge is to systematically inverse design photonic structures in targeted symmetry settings that realize the required isolated degeneracies at lower dielectric contrasts~\cite{vaidya2021point, kim2023automated, ghorashi2026symbolic, slobozhanyuk2017three, lu2013weyl}. Second, this mechanism opens the possibility of designing photonic environments to control radiative coupling from nanocavities, rather than relying on bandgap engineering. Related ideas have been explored in the context of Dirac-point cavities~\cite{dirac1, diracexp1, diracexp2, diracexp3, medina2023corner, wehling2007local} and symmetry-protected bound states in the continuum~\cite{vaidya2021point, cerjan2019bound, cerjan2021observation, cerjan2025lines}, where engineering the surrounding dispersion and symmetry landscape enables high-Q confinement in PhC fibers and slabs without full gaps. Prior studies of photonic Weyl media have also demonstrated weak confinement at defects near topologically-protected linear degeneracies~\cite{garcia2019tunable, ying2019extended}. Furthermore, confinement of this kind in the electronic context is also an interesting direction~\cite{thinel2024electronic, benalcazar2020bound, ghorashi2024highly, wehling2007local}. More broadly, our work points towards a paradigm in which light trapping is achieved not by eliminating states globally, but by modifying both the spectral and symmetry properties of the photonic environment of the cavity.

\vskip 2ex
\paragraph{Acknowledgements} 
We thank Thomas Christensen and Aaron Welters for helpful discussions. This work was supported by the U.S. Army Research Office through the Institute for Soldier Nanotechnologies at MIT under Collaborative Agreement Number W911NF-23-2-0121. Manxi Shi acknowledges funding from the MIT Undergraduate Research Opportunities Program (UROP). The MIT SuperCloud and Lincoln Laboratory Supercomputing Center provided computing resources that contributed to the results reported in this work. Copyediting of this manuscript was performed in part using GPT-5.3 and GPT-5.4 models.

\paragraph{Data availability}
The data and code that support the findings of this study are available at GitHub: \url{https://github.com/AliGhorashiCMT/localized_defects_in_3d/tree/main}.

\bibliography{references}

\end{document}

%% file: main-setup.tex
\usepackage[utf8]{inputenc}
\usepackage[english]{babel}
\usepackage{hyperref}

\usepackage{microtype} 
\usepackage{xspace} 
\usepackage{adjustbox}

\usepackage{txfonts}  
\usepackage{txfontsb} 

\usepackage{bm} 

\usepackage{xcolor}
\usepackage[]{graphicx} 
\graphicspath{{figs/}}

\usepackage[]{booktabs}
\usepackage{array}
\usepackage{layouts}
\usepackage{multirow}

\usepackage{enumerate}
\usepackage[inline]{enumitem}

\usepackage{xr}
\makeatletter
\newcommand*{\addFileDependency}[1]{
  \typeout{(#1)}
  \@addtofilelist{#1}
  \IfFileExists{#1}{}{\typeout{No file #1.}}
}
\makeatother

\usepackage{hyperref}
\hypersetup{colorlinks,
	linkcolor={blue!75!black!80!yellow},
	citecolor={blue!75!black!80!yellow},
	urlcolor={blue!75!black!80!yellow}
}

\usepackage[capitalize,nameinlink]{cleveref}

\crefname{subequations}{Eqs.}{Eqs.} 
\Crefname{subequations}{Eqs.}{Eqs.}
\crefformat{subequations}{#2Eqs.~(#1)#3}
\Crefformat{subequations}{#2Eqs.~(#1)#3}
\crefname{page}{p.}{p.} 
\crefname{table}{Table}{Tables}
\crefname{figure}{Figure}{Figures}
\crefname{section}{Section}{Sections}

\usepackage{placeins}

\usepackage{siunitx}
\DeclareSIUnit[number-unit-product = ]\percent{\char`\%} 

\usepackage[centering,hmargin=18mm,tmargin=29.4mm,bmargin=24mm]{geometry}

\usepackage{soul}

\usepackage{textcomp} 
\usepackage{xifthen}
\usepackage{etoolbox}
\newboolean{togglecomments}
\newboolean{toggletodos}
\newboolean{togglechanges}

\setboolean{togglecomments}{true}
\setboolean{toggletodos}{true}
\setboolean{togglechanges}{false} 

\newcommand{\textblacksquare}{$\blacksquare$}
\newcommand{\todo}[1]{\ifbool{toggletodos}%
	{\textcolor{green!60!black}{\small\textsf{{}\textsuperscript{\textsc{\textsf{todo}}}}[\ignorespaces#1]}} 
	{}}     
\newcommand{\comment}[2]{\ifbool{togglecomments}%
		{\textcolor{blue!70!black}{\small\sf\textsuperscript{\textsc{\textsf{\ignorespaces#1}}}[\ignorespaces#2]}} 
		{}}     
\newcommand{\swap}[2]{\ifbool{togglechanges}
	{\ignorespaces#2}  
	{\textcolor{red!70!black}{[\ignorespaces#1]}\textrightarrow{}\textcolor{green!50!black}{[\ignorespaces#2]}}}
\newcommand{\remove}[1]{\ifbool{togglechanges}
	{}    
	{\textcolor{blue}{\ignorespaces#1}}}
\newcommand{\inset}[1]{\ifbool{togglechanges}
	{\ignorespaces#1}  
	{\textcolor{green!50!black}{\ignorespaces#1}}}

\newcommand{\citeremind}[1]{%
	[\textcolor{blue!75!black!80!yellow}{\textblacksquare%
		\ifthenelse{\isempty{#1}}{}{\textsuperscript{\tiny\textsf{\ignorespaces#1}}}%
	}]\xspace}

\input{commands.tex}

\makeatletter
\newcommand{\raisemath}[1]{\mathpalette{\raisem@th{#1}}}
\newcommand{\raisem@th}[3]{\raisebox{#1}{$#2#3$}}
\makeatother

\renewcommand{\paragraph}[1]{\vskip 1ex\noindent\textbf{#1.}~}

\usepackage{braket}
\usepackage[eulergreek]{sansmath}
\makeatletter
\renewcommand\@make@capt@title[2]{%
    \@ifx@empty\float@link{\@firstofone}{\expandafter\href\expandafter{\float@link}}%
    \sisetup{math-sf=\textsf}%
    \sansmath\sffamily\textbf{#1\@caption@fignum@sep}#2 
}%

\makeatother


%% file: commands.tex
\newcommand{\appropto}{\mathrel{\vcenter{
			\offinterlineskip\halign{\hfil$##$\cr
				\propto\cr\noalign{\kern.2pt}\sim\cr\noalign{\kern-2.5pt}}}}}

\DeclareFontFamily{U}{mathx}{\hyphenchar\font45}
\DeclareFontShape{U}{mathx}{m}{n}{<5> <6> <7> <8> <9> <10>
                                  <10.95> <12> <14.4> <17.28> <20.74> <24.88>
                                  mathx10}{}
\DeclareSymbolFont{mathx}{U}{mathx}{m}{n}
\DeclareFontSubstitution{U}{mathx}{m}{n}